\documentclass{JHEP3} 

\usepackage{epsfig}
\usepackage{graphicx}
\usepackage{cite}
\usepackage[normalem]{ulem}
\usepackage{graphics}
\usepackage{amsmath}
\usepackage{amsthm}
\usepackage{amssymb}
\usepackage{enumerate}
\usepackage{bm,longtable,enumerate}
\usepackage{amsmath,amssymb}

\newcommand{\ol}{\overline}

\def\tr{\mathop{\rm tr}\nolimits}

\newcommand{\AD}[1]{$\ol{\mbox{D~\,}}\!\!\!$#1}

\newcommand{\eqn}{\begin{eqnarray}}
\newcommand{\eqnx}{\end{eqnarray}}

\def\beq{\begin{equation}}
\def\eeq{\end{equation}}
\def\beqa{\begin{eqnarray}}
\def\eeqa{\end{eqnarray}}
\def\ss{\scriptscriptstyle}

\def\matt[#1,#2,#3,#4]{\left(%
\begin{array}{cc} #1 & #2 \\ #3 & #4 \end{array} \right)}

\def\v2#1{\vv2[#1]}
\def\vv2[#1,#2]{\left(%
{#1 \atop #2}\right)}

\title{Deep inelastic scattering for vector mesons in holographic D4-D8 model}

\author{C.  A.  Ballon Bayona$^{\dag}$, Henrique Boschi-Filho$^{\ddag,a}$,
Nelson R. F. Braga$^{\ddag,b}$and Marcus A. C. Torres$^{\ddag,c}$
\\
$^\dag$Centro Brasileiro de Pesquisas F\'{i}sicas, Rua Dr. Xavier Sigaud 150,\\
Urca, 22290-180, Rio de Janeiro, RJ, Brazil\\
e-mail: ballon@cbpf.br
\\
\\
$^{\ddag}$Instituto de F\'{i}sica, Universidade Federal do Rio de Janeiro,\\
Caixa Postal 68528, RJ 21941-972, Brazil\\
$^{a}$e-mail: boschi@if.ufrj.br;
$^{b}$e-mail: braga@if.ufrj.br;
$^{c}$e-mail: mtorres@if.ufrj.br}

\abstract{We study deep inelastic scattering for  vector and axial vector mesons in the holographic D4-D8 brane model. We consider tree level contributions with one particle in the final hadronic state. We obtain the unpolarized structure functions $F_1$ and $F_2$ for the $\rho$ and $a_1$ mesons for $ q^2 < 80{\,\rm GeV}^2$ and $0.2 < x < 1\,$. We find that the ratio $F_2/(2xF_1)$ is approximately equal to one for some ranges of $x$ and $q^2$, satisfying the Callan-Gross relation.}

\preprint{}

 
\begin{document}

\section{Introduction}

The AdS/CFT correspondence inspired many interesting holographic models to describe non-perturbative aspects of strong interactions. This correspondence relates, in particular, string theory in $AdS_5 \times S^5 $ space to  ${\cal N} = 4$ Yang Mills $SU(N)$ theory with large $N$ in four dimensions \cite{Maldacena:1997re,Gubser:1998bc,Witten:1998qj}.

One of the first holographic models started with the idea of introducing an infrared cut off in the energy, related to some localization in the AdS bulk. Using this approach, the high energy scaling of the amplitude for glueball scattering at fixed angles was found in \cite{Polchinski:2001tt} (see also \cite{BoschiFilho:2002zs,Andreev:2002aw}).
This scaling was observed experimentally for other hadrons and was also obtained from  QCD 
\cite{QCD1,BRO}.  Glueball masses were calculated in ref. 
\cite{BoschiFilho:2002vd,BoschiFilho:2002ta} considering boundary conditions in AdS slice. This model is known as the hard wall model and motivated other holographic models for strong interactions that are usually called AdS/QCD models.

A simple way to describe flavour degrees of freedom in gauge/string duality consists on the inclusion of $N_f$ D7 brane probes in the D3 brane space. Then, one has open strings with an endpoint  on a D3 and the other on a D7 brane whose excitations are in the fundamental representation of the gauge group $SU(N_c)$, like quarks. On the other hand, fields related to open strings with both ends attached to the D3 branes are in the adjoint representation of the $SU(N_c)$ gauge group, like gluons.  This is known as the D3-D7 brane model, where mesons are described by strings with both endpoints on D7 branes \cite{Karch:2002sh,Kruczenski:2003be,Hong:2003jm}. 
The corresponding masses were calculated in \cite{Kruczenski:2003be}. 
For a review see \cite{Erdmenger:2007cm}.

A more sophisticated model that incorporates chiral symmetry breaking is the D4-D8 model proposed in \cite{Sakai:2004cn,Sakai:2005yt} (see also \cite{Hata:2007mb,Hashimoto:2008zw,Hashimoto:2009ys}). This model consists on the intersection of $N_c$ D4-branes and $N_f$ D8-\AD8 pairs of branes in type IIA string theory in the limit $N_f \ll N_c$. This corresponds to a probe limit for the D8-\AD8 branes in the D4 brane geometry. $N_c$ and $N_f$ are the color and flavour numbers of strongly coupled $SU(N_c)$ theory with massless quarks. The chiral symmetry breaking is realized geometrically through the merging of the D8-\AD8 branes.

Holographic models for QCD have also been used to study the interaction between photons and hadrons. 
Electromagnetic form factors have been calculated in these models, for example in \cite{Grigoryan:2007vg,Grigoryan:2007my,Brodsky:2007hb,Kwee:2007nq,RodriguezGomez:2008zp,Bayona:2009ar}. 
Some of these models, in particular the D4-D8 model, present the fundamental property that the interaction of hadrons with photons is mediated by vector mesons. This property is observed experimentally and is known as vector meson dominance \cite{GellMann:1961tg,Sakurai_69}.

Deep inelastic scattering (DIS) is an important tool to probe the internal structure of hadrons. 
From DIS one can calculate structure functions that are related to the distribution of partons inside hadrons. A detailed description of DIS using gauge/gravity duality was formulated 
in \cite{Polchinski:2002jw} in the hard wall model. 
For other works on DIS using gauge/string duality see, for instance, 
\cite{Hatta:2007he,BallonBayona:2007qr,BallonBayona:2007rs,Cornalba:2008sp,Pire:2008zf,
Albacete:2008ze,BallonBayona:2008zi,Yoshida:2009dw,Gao:2009ze,Hatta:2009ra,Avsar:2009xf,
Cornalba:2009ax,Bayona:2009qe,Cornalba:2010vk,Betemps:2010ij,Gao:2010qk,Kovchegov:2010uk}.

In this article we use the D4-D8 brane model to study DIS of unpolarized vector and axial-vector mesons. These particles appear, in this model, as an infinite tower of states. We take as initial states of the DIS process the lowest states corresponding to the $\rho$ vector meson and the $a_1$ axial-vector meson. We work at tree level and consider final states composed of just one vector or axial-vector meson, respectively. This corresponds to calculate the lowest order contribution to DIS structure functions. 
It is important to remark that the D4-D8 model was conceived as a phenomenological model for QCD at low energies ($\sim 1$ GeV). However, there are indications that this model can also reproduce QCD properties at high energies, like the behavior of electromagnetic form factors for vector and axial-vector mesons with large momentum transfer $q^2$  \cite{Bayona:2009ar}.

We calculate numerically the structure functions $F_1$ and $F_2$, and investigate their behavior with the momemtum transfer $q^2$ and the Bjorken parameter $x$ in the range of  
$ q^2~<~80{\,\rm GeV}^2$ and $0.2 < x < 1\,$. 
We also investigate if the Callan-Gross relation $F_2/(2xF_1)=1$, obtained in the parton model, is satisfied in the D4-D8 model. We find that this relation is approximately valid in the range 
of $0.4 < x < 0.6$ and momentum transfer $ 10{\,\rm GeV}^2~<~q^2~<~80{\,\rm GeV}^2$.

%%%%%%%%%%%%%%%%%%%%%%%%%%%%%%%%%%%%%%%%%%%%%%%%%%%%%%%%%%%%%%%%%%%%%%%%%

\section{Deep inelastic scattering}

Deep inelastic scattering consists of the scattering of a lepton on a hadron. Here we consider the case of vector and axial-vector mesons. 
The lepton produces a virtual photon of momentum $q^\mu$ which interacts with the hadron of momentum $P^\mu $. The final hadronic state is represented by $X$ with momentum $P_X^\mu$ 
(see Fig.~1). One can parametrize this process using as dynamical variables the photon virtuality $q^2$ and the Bjorken parameter $x \equiv  -q^2 /2P\cdot q \,$. Deep inelastic scattering corresponds to the large  $q^2$ limit, with $x$ fixed. For a review of DIS see \cite{Manohar:1992tz}. 

%%%%%%%%%%%%%%%%%%%%%%%%%%%%%%%%%%%%%%%%%%%%%%%%%%%%%%%%%%
%%%%%%%%%%%%%%%%%%%%%%%%%%%%%%%%%%%%%%%%%%%%%%%%%%%%%%%%%%
%%% figura do espalhamento profundamente inelastico   %%%%
%%%%%%%%%%%%%%%%%%%%%%%%%%%%%%%%%%%%%%%%%%%%%%%%%%%%%%%%%%
%%%%%%%%%%%%%%%%%%%%%%%%%%%%%%%%%%%%%%%%%%%%%%%%%%%%%%%%%%
\begin{figure}\begin{center}
\setlength{\unitlength}{0.1in}
\vskip 3.cm
\begin{picture}(0,0)(15,0)
\rm
%%%%%%%%%%%%%%%%%%% Lepton %%%%%%%%%%%%%%%%%%%%%%
\thicklines
\put(1,14.5){$\ell$}
\put(3,15){\line(2,-1){7}}
\put(3,15){\vector(2,-1){4}}
\put(18,14.5){$\ell$}
\put(17,15){\line(-2,-1){7}}
\put(10,11.5){\vector(2,1){4.2}}
%%%%%%%%%%%%%%%%%%%  Foton  %%%%%%%%%%%%%%%%%%%%%%%
\put(9.5,8){$q$}
\bezier{300}(10,11.5)(10.2,10.7)(11,10.5)
\bezier{300}(11,10.5)(11.8,10.3)(12,9.5)
\bezier{300}(12,9.5)(12.2,8.7)(13,8.5)
\bezier{300}(13,8.5)(13.8,8.3)(14,7.5)
%%%%%%%%%%%%%%%%%%   Proton  %%%%%%%%%%%%%%%%%%%%%%%
\put(0,-2){$P$}
\put(3,0){\line(2,1){10.5}}
\put(3,0){\vector(2,1){6}}
%%%%%%%%%%%%%%%%%% Interaction %%%%%%%%%%%%%%%%%%%%%
\put(16,6){\circle{5}}
%%%%%%%%%%%%%%%%%%   Hadrons %%%%%%%%%%%%%%%%%%%%%%%
\put(27,-2){$X$}
\put(18.5,5.5){\line(3,-1){8}}
\put(18.3,5){\line(2,-1){8}}
\put(18,4.5){\line(3,-2){7.5}}
\put(17.5,3.8){\line(1,-1){6}}
%%%%%%%%%%%%%%%%%%%%%%%%%%%%%%%%%%%%%%%%%%%%%%%%%%%%%
\end{picture}
\vskip 1.cm
\parbox{4.1 in}{\caption{} Illustrative diagram for a deep inelastic scattering. A lepton $\ell$ exchanges a virtual photon with a hadron of momentum $P$.}
\end{center}
\end{figure}
\vskip .5cm
%%%%%%%%%%%%%%%%%%%%%%%%%%%%%%%%%%%%%%%%%%%%%%%%%%%%%%
%%%%%%%%%%%%%%%%%%%%%%%%%%%%%%%%%%%%%%%%%%%%%%%%%%%%%%

The differential cross section of the process can be determined from the 
hadronic tensor, defined as 
\begin{equation}
W^{\mu\nu} \, = i \, \int d^4y\, e^{iq\cdot y} \langle P \vert \, \Big[ J^\mu (y) , J^\nu (0) \Big] 
\, \vert P \rangle \,,
 \label{HadronicTensor}
\end{equation}

\noindent where $ J^\mu(y)$ is the electromagnetic current of the meson. This tensor can be decomposed, in the unpolarized case, into the  structure functions $F_1 (x,q^2) $ and $F_2 (x,q^2) $ as \cite{Manohar:1992tz}
\begin{equation}
W^{\mu\nu} \, = \, F_1 (x,q^2)  \Big( \eta^{\mu\nu} \,-\, \frac{q^\mu q^\nu}{q^2} \, \Big) 
\,+\,\frac{2x}{q^2} F_2 (x,q^2)  \Big( P^\mu \,+ \, \frac{q^\mu}{2x} \, \Big) 
\Big( P^\nu \,+ \, \frac{q^\nu}{2x} \, \Big)
\, ,  \label{Structure}
\end{equation}

\noindent where we use the Minkowski metric $\eta_{\mu\nu}={\rm diag}(-,+,+,+)$. 

As is well known, the cross section for the deep inelastic scattering is related to
the amplitude of forward Compton scattering. This amplitude is determined  by  the tensor 
\begin{equation}
T^{\mu\nu} \, = i \, \int d^4y e^{iq\cdot y} \langle P \vert \, {\cal T} \Big(  J^\mu (y) J^\nu (0) \Big)  \, \vert P \rangle\,,
 \label{forwardamplitude}
\end{equation}

\noindent which can be decomposed in the same way as in eq. (\ref{Structure}), but with 
structure functions ${\tilde F}_1 (x,q^2)  $ and ${\tilde F}_2 (x,q^2) $.  
The optical theorem implies that~\cite{Manohar:1992tz}
\begin{equation}
\label{optical}
F_{1,2} (x,q^2) \equiv 2 \pi \,{\rm Im }\,{\tilde F}_{1,2} (x,q^2)\,.
\end{equation}

Below, we will investigate DIS for vector and axial-vector mesons in the holographic D4-D8 model. In this model the masses and couplings of vector mesons arise from the Kaluza-Klein expansion of gauge fields in a 5d effective action. We will see that the imaginary part of the Compton forward tensor 
eq. (\ref{forwardamplitude}) can be obtained from these couplings and masses. 

%%%%%%%%%%%%%%%%%%%%%%%%%%%%%%%%%%%%%%%%%%%%%%%%%%%%%%%%

%%%%%%%%%%%%%%%%%%%%%%%%%%%%%%%%%%%%%%%%%%%%%%%%%%%%%%%% 

\section{Vector mesons in D4-D8 Model}

The Sakai-Sugimoto D4-D8 model is built adding $N_f$ pairs of \mbox{D8}  and \AD8 probe branes in the spacetime background formed by the presence of $N_c$ \mbox{D4} branes. 
The background forces the branes D8 and \AD8  to merge in a single stack of $N_f$ D8 branes, breaking chiral symmetry $U(N_f)_L\times U(N_f)_R$ into a remaining $U(N_f)$. 

In this model, mesons of the dual gauge theory appear as states of open strings connecting the D8-\AD8 branes. These states correspond to fluctuations of the D8 branes solutions in the D4 background. In particular, vector and axial-vector mesons are described by  $U(N_f)$ gauge field fluctuations. 

The dynamics of these fluctuations is given by the Dirac Born Infeld action of the D8-\AD8 embedding, which leads to the five dimensional effective action 
\cite{Sakai:2005yt}
\beq
S_{eff}\,=\, \kappa \int d^4 x \int d \tilde z \, \tr  \, \left [ \frac{1}{2} (K (\tilde z))^{-1/3} \eta^{\mu \lambda} \eta^{\nu \rho} F_{\lambda \rho} F_{\mu \nu} + M^2_{\ss{KK}} K(\tilde z) \eta^{\mu \nu} F_{\mu \tilde z} F_{\nu \tilde z}\right ] \label{effectivegaugeaction}
\eeq
where
%\footnote{Note that our coordinate $\tilde z$ is related to the coordinate $z$ of ref. \cite{Sakai:2005yt} by $\tilde z=z/U_{\ss{KK}}$.} 
$\tilde z$ is a dimensionless variable with range $(-\infty,+\infty)$ that connects the original left and right chiral sectors,  
\beq
K(\tilde z) \, \equiv \, 1 + \tilde z^2 \quad , \quad \kappa \,=\, \frac{g_{YM}^2 N_c^2}{ 216 \pi^3} \,,
\eeq
and $M_{\ss{KK}}$ is a mass parameter related to the D4 brane background. 
We set $M_{\ss{KK}}=946$~MeV to fit the mass of $\rho$(770) meson. 

Expanding the gauge fields, in the gauge $A_{\tilde z}=0$, as \cite{Sakai:2005yt}:
\beqa
A_{\mu} (x,\tilde z) &\,=\,&  \hat {\cal V}_\mu (x)  + \hat {\cal A}_\mu (x)  \psi_0 (\tilde z)  \,+\, \sum_{n=1}^{\infty} v_\mu^n (x) \psi_{2n-1} (\tilde z) \,+\,  \sum_{n=1}^{\infty} a_\mu^n (x) \psi_{2n} (\tilde z)    \quad , \quad \label{gaugefieldexpansion}
\eeqa
where $\psi_0 (\tilde z)\equiv (2/\pi ) \arctan{\tilde z}$ and 
\beqa
\hat {\cal V}_\mu (x)&\,=\,& \frac12  e^{- \frac{ i \Pi(x)}{f_\pi}}\left[ A_{L\mu}(x) + \partial_\mu \right] e^{\frac{ i \Pi(x)}{f_\pi}} \,+\, \frac12  e^{ \frac{ i \Pi(x)}{f_\pi}}\left[ A_{R \mu} (x) + \partial_\mu \right] e^{\frac{- i \Pi(x)}{f_\pi}}\nonumber \\
\hat {\cal A}_\mu (x) &\,=\,& \frac12  e^{- \frac{ i \Pi(x)}{f_\pi}}\left[ A_{L\mu} (x)+ \partial_\mu \right] e^{\frac{ i \Pi(x)}{f_\pi}} \,-\, \frac12  e^{ \frac{ i \Pi(x)}{f_\pi}}\left[ A_{R \mu} (x) + \partial_\mu \right] e^{\frac{- i \Pi(x)}{f_\pi}} \,.
\eeqa
The field $\Pi(x)$ is interpreted as the pion field. The modes $\psi_n$  satisfy the following conditions 
\beqa
\kappa \int d \tilde z (K( \tilde z))^{-1/3} \psi_r (\tilde z) \psi_s (\tilde z) &\,=\,&  \delta_{rs} \, \, ,  \label{vectormesonnorm} \\
- (K(\tilde z) )^{1/3} \partial_{\tilde z} \left[ K(\tilde z) \partial_{\tilde z} \psi_r (\tilde z) \right] &\,=\,& \lambda_r \,  
\psi_r (\tilde z) \, \label{vectormesoneq},
\eeqa
where $r,s$ are positive integers. 

In order to represent vector and axial-vector mesons, one performs the following field redefinitions 
\beqa
\tilde v_\mu^n &\,=\,& v_\mu^n + \frac{g_{v^n}}{M_{v^n}^2} {\cal V}_\mu  \quad \, \, \, \, , \quad 
\tilde a_\mu^n \,=\, a_\mu^n + \frac{g_{a^n}}{M_{a^n}^2} {\cal A}_\mu  \, ,  \\
 {\cal V}_\mu &\,=\,& \frac12 (A_{L\mu} + A_{R\mu})\quad , \quad {\cal A}_\mu \,=\, \frac12 (A_{L\mu} - A_{R\mu}) \, ,
\eeqa
introducing the constants
\beqa
M_{v^n}^2 &\,=\,& \lambda_{2n-1} M^2_{\ss{KK}} \qquad \, \, , \quad M_{a^n}^2 \,=\, \lambda_{2n} M^2_{\ss{KK}} \, , \label{masses}\\
 g_{v^n} &\,=\,& \kappa  \, M_{v^n}^2 \int d \tilde z \,K(\tilde z)^{-1/3} \psi_{2n-1}(\tilde z) \, , \label{decayvn}\\
g_{a^n} &\,=\,& \kappa \, M_{a^n}^2  \int d \tilde z \, K(\tilde z)^{-1/3} \psi_{2n}(\tilde z) \psi_0(\tilde z) \label{decayan} \, .
\eeqa

This way, one obtains the 4d effective Lagrangian 
\beqa
{\cal L}^{4d}_{eff} &\,=\,& \frac12 \tr \left(\partial_\mu \tilde v_\nu^n - \partial_\nu \tilde v_\mu^n \right)^2 + \frac12 \tr  \left(\partial_\mu \tilde a_\nu^n - \partial_\nu \tilde a_\mu^n \right)^2 + \tr  \left(i \partial_\mu \Pi + f_\pi {\cal A}_\mu \right)^2 \nonumber \\
&\,+\,& M_{v^n}^2 \tr  \left(\tilde v_\mu^n - \frac{g_{v^n}}{M_{v^n}^2} {\cal V}_\mu \right)^2 + M_{a^n}^2 \tr  \left(\tilde a_\mu^n - \frac{g_{a^n}}{M_{a^n}^2} {\cal A}_\mu \right)^2 \,+\, \sum_{j \ge 3} {\cal L}_j \,  \label{fourdimensionalefflag}
\eeqa
where ${\cal L}_j$ represent the interaction terms of order $j$ in the fields and divergent terms were disregarded.

The massive fields $\tilde v_\mu^n$, $\tilde a_\mu^n$ represent vector and axial-vector mesons, respectively. The decay constant $g_{v^n}$  couples the vector mesons $\tilde v_\mu^n$ to the photon 
${\cal V}_\mu$. 
The other decay constant $g_{a^n}$ couples the axial-vector meson $\tilde a_\mu^n$ to a massless axial vector field ${\cal A}_\mu$, that is turned off, setting $A_{L\mu} = A_{R\mu}$. 
Note that $g_{v^n}$ is the only interaction between photons and mesons, which implies that vector meson dominance is realized in the D4-D8 model.

In order to calculate structure functions, we need the coupling among three vector mesons and also among one vector and two axial-vector mesons. These couplings appear in the following contribution of the 4d interaction lagrangian 
${\cal L}_j$ 
\beqa
 {\cal L}^{4d}_{int} &\,=\,&  \tr \Big\{  \left(\partial^\mu {\tilde v}^{\nu\, n} - \partial^\nu {\tilde v}^{\mu \, n} \right) 
\Big( g_{v^n v^\ell v^m} [{\tilde v}_\mu^\ell, {\tilde v}_\nu^m]  
+ g_{v^n a^\ell a^m} [{\tilde a}_\mu^\ell, {\tilde a}_\nu^m]   \Big) \cr 
&& + g_{v^\ell a^m a^n} \left(\partial^\mu {\tilde a}^{\nu\, n} - \partial^\nu {\tilde a}^{\mu \, n} \right) 
 \Big( [{\tilde v}_\mu^\ell, {\tilde a}_\nu^m] - [{\tilde v}_\nu^\ell, {\tilde a}_\mu^m] \Big)  \Big\}\,, \label{IntLagrangian}
\eeqa

\noindent where the three-meson coupling constants are given by 
\beqa
g_{v^n v^\ell v^m} &\,=\,& \kappa \,  \int d \tilde z \,K(\tilde z)^{-1/3} 
\psi_{2n-1}(\tilde z) \psi_{2\ell -1}(\tilde z) \psi_{2m-1}(\tilde z) \label{couplingvvv}
\, ,\\
g_{v^\ell a^m a^n} &\,=\,& \kappa  \int d \tilde z \, K(\tilde z)^{-1/3} 
\psi_{2\ell -1}(\tilde z) \psi_{2m}(\tilde z)\psi_{2n}(\tilde z)  \, .\label{couplingvaa}
\eeqa

\begin{table}
\bigskip\centerline{\begin{array}[b]{|c|c|c|c|c|c|c|c|}
\hline
 &&&&&&&\\
n&\frac{M_{v^n}^2}{M_{KK}^2}&\frac{g_{v^n}}{\sqrt{\kappa}M_{KK}^2}&\sqrt{\kappa}g_{v^nv^1v^1}& \sqrt{\kappa}g_{v^nv^1v^2} & \sqrt{\kappa}g_{v^nv^1v^3} &\sqrt{\kappa}g_{v^nv^1v^4}& \sqrt{\kappa}g_{v^nv^1v^5}\\
&&&&&&& \\
\hline\hline
1  & 0.6693     & 2.109 & 0.4466 &-0.1465 &0.0184& -3.689{\ss\times} 10^{\ss -4}& 2.696{\ss\times} 10^{\ss -4} \\
2  & 2.874      & 9.108&  -0.1465 &   0.2687 &-0.1477& 0.0237& -1.921{\ss\times} 10^{\ss -4} \\
3 & 6.591       & 20.80 &0.0184 &  -0.1477 & 0.2522& -0.1487& 0.0255  \\
4 & 11.80       & 37.15&  -3.689{\ss\times} 10^{\ss -4}&0.0237& -0.1487& 0.2473& -0.1491 \\
5 & 18.49       & 58.17  &  2.695{\ss\times} 10^{\ss -4}&  - 1.921 {\ss\times}10^{\ss -4}&0.0255& -0.1491& 0.2451\\
6 & 26.67       & 83.83 &  3.078{\ss\times} 10^{\ss -5}& 4.469 {\ss\times}10^{\ss -4}&-7.647{\ss\times}10^{\ss -5}& 0.0263& -0.1493  \\
7 & 36.34       & 114.2& 1.857{\ss\times} 10^{\ss-5}&   7.560 {\ss\times}10^{\ss -5} &5.374{\ss\times}10^{\ss -4}& -1.033{\ss\times}10^{\ss -5}& 0.0268\\
8 & 47.49       & 149.1 &  6.996{\ss\times} 10^{\ss -6}&4.080 {\ss\times}10^{\ss -5} &1.044{\ss\times}10^{\ss -4}& 5.888{\ss\times}10^{\ss -4}& 2.991{\ss\times}10^{\ss -5}\\
9& 60.14        & 188.7  & 3.508{\ss\times} 10^{\ss -6}&1.723 {\ss\times}10^{\ss -5}& 5.579{\ss\times}10^{\ss -5}& 1.230{\ss\times}10^{\ss -4}& 6.204{\ss\times}10^{\ss -4}  \\ \hline
\end{array}}\caption{Dimensionless squared masses and coupling constants for vector mesons.}
\label{vectortableconstants}
\end{table}

\begin{table}
\bigskip\centerline{\begin{array}[b]{|c|c|c|c|c|c|}
\hline
 &&&&&\\
n&\sqrt{\kappa}g_{v^nv^1v^6}& \sqrt{\kappa}g_{v^nv^1v^7} & \sqrt{\kappa}g_{v^nv^1v^8} &\sqrt{\kappa}g_{v^nv^1v^9}& \sqrt{\kappa}g_{v^nv^1v^{10}}\\
&&&& &\\ 
\hline\hline
1  & 3.077{\ss\times} 10^{\ss -5}&1.857{\ss\times} 10^{\ss -5}& 6.996{\ss\times} 10^{\ss -6}& 3.506{\ss\times} 10^{\ss -6}& 1.809{\ss\times} 10^{\ss -6} \\
2  & 4.469{\ss\times} 10^{\ss -4}& 7.560{\ss\times} 10^{\ss -5}& 4.080{\ss\times} 10^{\ss -5}& 1.723{\ss\times} 10^{\ss -5}& 8.922{\ss\times} 10^{\ss -6} \\
3 & -7.650{\ss\times} 10^{\ss -5}& 5.375{\ss\times} 10^{\ss -4}& 1.045{\ss\times} 10^{\ss -4}& 5.590{\ss\times} 10^{\ss -5}& 2.489{\ss\times} 10^{\ss -5}  \\
4 &0.0263& -1.055{\ss\times} 10^{\ss -5}& 5.888{\ss\times} 10^{\ss -4}& 1.229{\ss\times} 10^{\ss -4}& 6.599{\ss\times} 10^{\ss -5}\\
5 &-0.1493& 0.0268& 2.954{\ss\times} 10^{\ss -5}&6.205{\ss\times} 10^{\ss -4}& 1.350{\ss\times} 10^{\ss -4}\\
6 &0.2439& -0.1494& 0.0271& 5.552{\ss\times} 10^{\ss -5}& 6.413{\ss\times} 10^{\ss -4}\\
7 &-0.1494& 0.2432& -0.1495& 0.0272& 7.318{\ss\times} 10^{\ss -5}\\
8 &0.0271& -0.1495& 0.2428& -0.1495& 0.0273\\
9& 5.574{\ss\times} 10^{\ss -5}& 0.0272& -0.1496& 0.2425& -0.1496 \\ \hline
\end{array}}\caption{More triple vertex coupling constants for vector mesons.}
\label{vectortableconstants2}
\end{table}

\begin{table}
\bigskip\centerline{\begin{array}[b]{|c|c|c|c|c|c|c|}
\hline
 & & & &&& \\
n & \frac{M_{a^n}^2}{M_{KK}^2}  & \sqrt{\kappa}g_{v^na^1a^1} & \sqrt{\kappa}g_{v^na^1a^2}& \sqrt{\kappa}g_{v^na^1a^3} & \sqrt{\kappa}g_{v^na^1a^4} & \sqrt{\kappa}g_{v^na^1a^5} \\[0.5ex]
&&&&&& \\
\hline\hline
1& 1.569 &   0.2865 & -0.1453&0.02180& -2.774 {\ss\times}10^{\ss -4}&3.737 {\ss\times}10^{\ss -4}\\
2& 4.546&   0.1475 & 0.1345  & -0.1460& 0.0287& 7.508 {\ss\times}10^{\ss -5}\\
3& 9.008  &  -0.1438   &  0.1248 &  0.1205& -0.1469& 0.03123\\
4 & 14.96&   0.0262  & -0.1465  & 0.1179& 0.1162& -0.1473\\
5 & 22.39   &   8.738 {\ss\times}10^{\ss -5}  & 0.0302  & -0.1471& 0.1150& 0.1142\\
6& 31.32 &   5.312 {\ss\times}10^{\ss -4}     &  1.995  {\ss\times}10^{\ss -4}& 0.0319& -0.1474& 0.1136\\
7 & 41.73&   9.789 {\ss\times}10^{\ss -5} & 7.289  {\ss\times}10^{\ss -4}     &3.634 {\ss\times}10^{\ss -4}& 0.0328& -0.1475\\
8 & 53.63&   5.144{\ss\times}10^{\ss -5}  & 1.622 {\ss\times}10^{\ss -4}  &8.395 {\ss\times}10^{\ss -4}& 4.611 {\ss\times}10^{\ss -4}& 0.0333\\
9& 67.02 &   2.212{\ss\times}10^{\ss -5}   & 8.421 {\ss\times}10^{\ss -5} & 2.037 {\ss\times}10^{\ss -4}&9.069 {\ss\times}10^{\ss -4}&5.232 {\ss\times}10^{\ss -4}\\ \hline
\end{array}}\caption{Dimensionless squared masses and triple vertex coupling constants for axial-vector mesons.}
\label{axialtableconstants}
\end{table}

\begin{table}
\bigskip\centerline{\begin{array}[b]{|c|c|c|c|c|c|}
\hline
 & & &&& \\
n &\sqrt{\kappa}g_{v^na^1a^6} & \sqrt{\kappa}g_{v^na^1a^7}& \sqrt{\kappa}g_{v^na^1a^8} & \sqrt{\kappa}g_{v^na^1a^9} & \sqrt{\kappa}g_{v^na^1a^{10}} \\[0.5ex]
&&&&& \\ 
\hline\hline
1&
 5.543{\ss\times} 10^{\ss -5}& 3.061{\ss\times} 10^{\ss -6}& 1.238{\ss\times} 10^{\ss -5}& 6.381{\ss\times} 10^{\ss -6}& 
  3.446{\ss\times} 10^{\ss -6}\\
2&6.456{\ss\times} 10^{\ss -4}& 1.335{\ss\times} 10^{\ss -4}& 6.918{\ss\times} 10^{\ss -5}& 3.094{\ss\times} 10^{\ss -5}& 1.652{\ss\times} 10^{\ss -5}
\\
3& 
 2.928{\ss\times} 10^{\ss -4}&7.916{\ss\times} 10^{\ss -4}& 1.849{\ss\times} 10^{\ss -4}& 9.625{\ss\times} 10^{\ss -5}&4.514{\ss\times} 10^{\ss -5}
\\
4 & 
 0.0324& 4.179{\ss\times} 10^{\ss -4}& 8.770{\ss\times} 10^{\ss -4}& 2.187{\ss\times} 10^{\ss -4}& 1.148{\ss\times} 10^{\ss -4}
\\
5 &
 -0.1474& 0.0331& 4.957{\ss\times} 10^{\ss -4}& 9.310{\ss\times} 10^{\ss -4}& 2.410{\ss\times} 10^{\ss -4}
\\
6& 
 0.1131& -0.1475& 0.0335& 5.458{\ss\times} 10^{\ss -4}& 9.667{\ss\times} 10^{\ss -4}\\
7 &
 0.1127& 0.1124& -0.1476& 0.0337& 5.805{\ss\times} 10^{\ss -4}
\\
8 & 
 -0.1476& 0.1122& 0.1120& -0.1476& 0.0339
\\
9&
 0.0336& -0.1476& 0.1119& 0.1117& -0.1477\\ \hline
\end{array}}\caption{More triple vertex coupling constants for axial-vector mesons.}
\label{axialtableconstants2}
\end{table}

In order to obtain the masses $\,M_{v^n}\,$, $\,M_{a^n}\,$ and the 
coupling constants $\,g_{v^n}\,$, $\,g_{v^n v^\ell v^m}\,$ and $\,g_{v^\ell a^m a^n}\,$ one has to calculate numerically the wave functions $\psi_{r}(\tilde z)$. 
We solved numerically the equations of motion for the vector and axial-vector modes using  the {\it shooting method}, following refs. \cite{Sakai:2004cn,Sakai:2005yt,Bayona:2009ar}. 
Some of these masses and couplings were calculated in these references. 
The determination of the structure functions (to lowest order) requires the calculation of many additional couplings among three vector and axial-vector mesons. We performed the numerical calculations of these couplings and list some results for the vector mesons in Tables 1 and 2, 
and for the axial-vector mesons in Tables 3 and 4. Note that the decay constants are positive and increase monotonically with $n$. The vector meson couplings $g_{v^nv^\ell v^m}$ and 
$g_{v^n a^\ell a^m}$ oscillate  with $n$ with decreasing absolute value. 

From these couplings one can calculate, for instance, the electromagnetic form factors that 
characterize the interaction of a vector meson with a photon. 
As shown in ref. \cite{Bayona:2009ar}, these form factors can be written as 
\begin{equation}
F_{ v^i v^j } (q^2) = \sum_n \frac{ g_{v^n } g_{v^i v^j v^n} }{ q^2 + m_n^2} 
\,, \label{formfactors}
\end{equation}

\noindent and a similar expression for axial-vector mesons:
\begin{equation}
F_{ a^i a^j } (q^2) = \sum_n \frac{ g_{v^n } g_{a^i a^j v^n} }{ q^2 + m_n^2} 
\,. \label{axialformfactors}
\end{equation}

%%%%%%%%%%%%%%%%%%%%%%%%%%%%%%%%%%%%%%%%%%%%%%%%%%%%%

%%%%%%%%%%%%%%%%%%%%%%%%%%%%%%%%%%%%%%%%%%%%%%%%%%%%%

\section{Structure functions of $\rho$ and $a_1$ mesons}

Here we will calculate the DIS structure functions for unpolarized vector mesons $\rho$ and  axial-vector mesons $a_1$. We will consider in this article the lowest order contribution with one vector or axial vector meson in the final state. 

\subsection{Analytical results}

 %%%%%%%%%%%%%%%%%%%%%%%%%%%%%%%%%%%%%%%%%% Figura %%%%%%%%%%%%%%%%%%%%%%%%%%
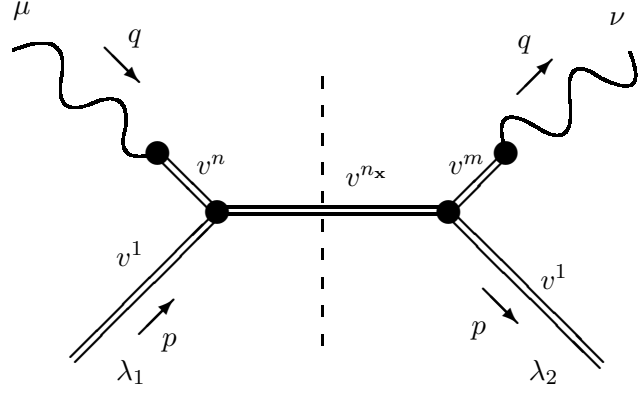
\begin{figure}
\begin{center}
\vskip 4cm
\begin{picture}(0,0)(20,0)
\setlength{\unitlength}{0.06in}
\rm
\thicklines 
%%%%%%%%%%%%%%%%%%%%%%%%%%%%%%%%%%%%%%%%% Mesons Vetoriais %%%%%%%%%%%%%%%%%%%%%%
\put(-3,11){$v^n$}
\put(-6,12){\line(1,-1){4.5}}
\put(-5.6,12.4){\line(1,-1){4.5}} 
\put(-6.4,12.8){\circle*{2}}
\put(-10,3){$v^1$}
\put(-6,-4){$p$}
\put(-10,-7){$\lambda_1$}
\put(-8,-3){\vector(1,1){3}}
\put(-1.2,7.7){\circle*{2}}
\put(10,10){$v^{n_{\bf x}}$}
\put(-14,-5){\line(1,1){13}}
\put(-13.6,-5.4){\line(1,1){13}}
\put(19,7.7){\circle*{2}}
\put(-1.2,8.1){\line(1,0){19.5}}
\put(-1.2,7.5){\line(1,0){19.5}}
\put(19,11){$v^m$}
\put(24,12.8){\circle*{2}}
\put(19,8.1){\line(1,1){4.5}}
\put(19.4,7.7){\line(1,1){4.5}} 
\put(19,8.1){\line(1,-1){13.5}}
\put(19.2,7){\line(1,-1){13}} 
\put(27,1){$v^1$}
\put(21,-3){$p$}
\put(22,1){\vector(1,-1){3}}
\put(26,-7){$\lambda_2$}
%%%%%%%%%%%%%%%%%%%%%%%%%%  Foton  inicial %%%%%%%%%%%%%%%%%%%%%
\put(-9,22.5){$q$}
\put(-11,22){\vector(1,-1){3}}
\put(-19,25){$\mu$}
\bezier{300}(-6.4,12.8)(-10.2,11.7)(-9.2,15)
\bezier{300}(-9.2,15)(-8.8,18.5)(-12,17.2)
\bezier{300}(-12,17.2)(-15.4,15.7)(-14.8,19.4)
\bezier{300}(-14.8,19.4)(-14,23.8)(-19,21.8)
%%%%%%%%%%%%%%%%%%%%%%%%%%  Foton  final %%%%%%%%%%%%%%%%%%%%%%%
\put(25,22){$q$}
\put(25,18){\vector(1,1){3}}
\put(33,24){$\nu$}
\bezier{300}(24,12.8)(22.5,17.7)(26.8,15)
\bezier{300}(26.8,15)(30,13)(29.6,17.2)
\bezier{300}(29.6,17.2)(29,21.7)(32.2,19.4)
\bezier{300}(32.2,19.4)(36.6,17.2)(34.8,21.6)
%%%%%%%%%%%%%%%%%%%%%%%%%%%%%%%%%%%%%%%%%%%%%%%%%%%%%%%%%%%%%%%%
\multiput(8,-4)(0,2.5){10}{\line(0,1){1}}
\end{picture}
\vskip 2.cm
\parbox{4.1 in}{\caption{Feynman diagram corresponding to the imaginary part of the vector meson Compton forward scattering amplitude. A similar diagram holds for the axial-vector meson replacing the lines $v^1, v^{n_{\bf x}}$ by $a^1, a^{n_{\bf x}}$.}}
\end{center}\label{Feynman}
\end{figure}
\vskip .5cm
%%%%%%%%%%%%%%%%%%%%%%%%%%%%%%%%%%%%%%%%%%%%%%%%%%%%%%%%%%%%%%%%%%%%%%%%%%%%%%%%%%

We start the discussion with the case of the $\rho$ vector meson represented by the lowest state 
$v^1$. The imaginary part of the Compton forward scattering amplitude associated with this process is represented in the Feynman diagram of Figure~2, which exhibits vector meson dominance. 
Using Feynman rules corresponding to this diagram, we find
\begin{eqnarray}
{\rm Im} T_{\mu\nu}&=&  \frac{1}{3} 
\sum_{\epsilon} \epsilon^{\lambda_1} \epsilon^{\lambda_2}
\sum_{n_{\bf x}} \left\{\sum_n g_{v^n }  g_{v^1 v^{n_{\bf x}} v^n} 
\frac{ \left[ \eta^{\mu \sigma_1} + \frac{q^{\mu} q^{\sigma_1}}{m_n^2}\right] }{q^2 + m_n^2} \right\} \cr\cr
&\times& \left\{ \sum_m g_{ v^m}g_{v^1 v^{n_{\bf x}} v^m} 
\frac{ \left[ \eta^{\nu \sigma_2} + \frac{q^{\nu} q^{\sigma_2}}{m_m^2}\right] }{q^2 + m_m^2} \right\} 
\cr\cr
&\times& f^{0ab} \left[ \eta_{\sigma_1 \lambda_1} (2q)_{\rho_1} + \eta_{ \lambda_1 \rho_1} (2p)_{\sigma_1} +
\eta_{\sigma_1 \rho_1} (-2q)_{\lambda_1} \right] \cr\cr
&\times& f^{0ab} \left[ \eta_{\sigma_2 \lambda_2} (2q)_{\rho_2} + \eta_{ \lambda_2 \rho_2} (2p)_{\sigma_2} +
\eta_{\sigma_2 \rho_2} (-2q)_{\lambda_2} \right] \cr\cr
 &\times &
\int \frac{d^4p_{_X}}{2\pi^4} (2\pi) \delta ( p_{_X}^2 + m_{n_{\bf x}}^2 ) 
\left[ \eta^{ \rho_1 \rho_2} + \frac{p_{_X}^{\rho_1} p_{_X}^{\rho_2}}{m_{n_{\bf x}}^2}\right] (2\pi)^4 
\delta^4 (p+q-p_{_X} )
\label{ImTmunu}
\end{eqnarray} 
  
\noindent where $\eta_{\mu\nu}$ is the Minkowski metric $(-+++)$, 
$p_X$ is the momentum of the intermediate mesonic state (final for the DIS process) with mass $m_{n_{\bf x}}$ satisfying the on-shell condition given by the delta term $\delta ( p_{_X}^2 + m_{n_{\bf x}}^2 )$. Note that the mass $m_{n_{\bf x}}$ has to coincide with one of the vector meson masses $m_j$ $(j=1, 2, 3, ...)$. The four-vectors $\epsilon^{\lambda_1}$ and $\epsilon^{\lambda_2}$ are the polarizations of the initial and final states of the forward scattering satisfying  
\begin{equation}
\sum_\epsilon \epsilon^{\lambda_1} \epsilon^{\lambda_2} = \eta^{\lambda_1 \lambda_2} +
\frac{ p^{\lambda_1} p^{\lambda_2} }{m_1^2} \,.
\end{equation}

In eq. (\ref{ImTmunu}), $f^{0ab}$ is the structure constant of the $U(N_f)$ flavor group. 
The photon is the gauge field of a $U(1)$ subgroup of $U(N_f)$ \cite{Sakai:2005yt}, 
here represented by the generator index~0. We consider the DIS process blind concerning the 
flavor group, and we sum the indices $a, b$ over all group generators.

One can rewrite the factors between braces in eq. (\ref{ImTmunu}), corresponding to vector mesons interacting with photons, in terms of the form factors $F_{v^1 v^{n_{\bf x}}} (q^2) $, defined in eq. (\ref{formfactors}), as 
\begin{equation}
\sum_n \frac{ g_{v^n } g_{v^1 v^{n_{\bf x}} v^n} }{ q^2 + m_n^2} 
\left[ \eta^{\mu \sigma_1} + \frac{q^{\mu} q^{\sigma_1}}{m_n^2}\right] = F_{ v^1 v^{n_{\bf x}} } (q^2) 
\left[ \eta^{\mu \sigma_1} - \frac{q^{\mu} q^{\sigma_1}}{q^2} \right] + F_{ v^1 v^{n_{\bf x}} } (0) 
\frac{q^{\mu} q^{\sigma_1}}{q^2}\,.
\end{equation}

\noindent The last term of this equation does not contribute to the amplitude given by eq. (\ref{ImTmunu}). Then, using the above results and the property $f^{0ab}f^{0ab}=N_f$ we find 
\begin{eqnarray}
{\rm Im} T_{\mu\nu}&=&   \frac{N_f}{3}  
\left[\eta^{\mu \sigma_1} - \frac{q^{\mu} q^{\sigma_1}}{q^2}\right] 
\left[\eta^{\nu \sigma_2} - \frac{q^{\nu} q^{\sigma_2}}{q^2}\right] 
\left[\eta^{\lambda_1 \lambda_2} + \frac{ p^{\lambda_1} p^{\lambda_2} }{m_1^2}\right]\cr\cr
&\times&\left[\eta^{ \rho_1 \rho_2} + \frac{(p+q)^{\rho_1} (p+q)^{\rho_2}}{s}\right] 
 \left[ \eta_{\sigma_1 \lambda_1} (2q)_{\rho_1} + \eta_{ \lambda_1 \rho_1} (2p)_{\sigma_1} +
\eta_{\sigma_1 \rho_1} (-2q)_{\lambda_1} \right] \cr\cr 
&\times& \left[ \eta_{\sigma_2 \lambda_2} (2q)_{\rho_2} + \eta_{ \lambda_2 \rho_2} (2p)_{\sigma_2} +
\eta_{\sigma_2 \rho_2} (-2q)_{\lambda_2} \right] \cr\cr
 &\times &
\sum_{n_{\bf x}} \left[F_{v^1 v^{n_{\bf x}}} (q^2)\right]^2
 (2\pi) \delta [ m_{n_{\bf x}}^2 -s ] 
\label{ImTmunu2}
\end{eqnarray} 

\noindent where $s=-(p+q)^2$. Note that, since $s$ is fixed by the initial momenta $p$ and $q$, the delta function selects only one value of 
$m_{n_{\bf x}}=m_{\bar n}=\sqrt{s}$, corresponding to a particular state $v^{\bar n}$.
Considering that the differences between the subsequent masses of the mesons are  small compared with their absolute value, one can approximate the sum over the delta functions by an integral 
\begin{eqnarray}
\sum_{n_{\bf x}} \delta [ m_{n_{\bf x}}^2 -s ] 
&\equiv& \sum_{n} \delta [ m_{n}^2 -m_{\bar n}^2 ] 
= \int dn 
\left[\left| \frac{\partial m_n^2}{\partial n} \right|\right]^{-1}
\delta(n-\bar n)\cr
&=&\left[\left| \frac{\partial m_n^2}{\partial n} \right|\right]_{n=\bar n}^{-1}\equiv f(\bar n)\,.
\end{eqnarray}

Then, the structure functions are
\begin{eqnarray}
{F}_1  &=&  \frac{16\pi^2 N_f}{3}  f(\bar n)  
 \Big[ F_{ v^1 v^{\bar n}}  (q^2) \Big]^2 q^2 \left[ 2 + \frac{q^2}{4x^2 m_1^2} + \frac{ q^2} {s x^2} ( x - \frac 12)^2 \right]
\cr
{F}_2  &=&  \frac{16\pi^2 N_f}{3}  f(\bar n) 
 \Big[ F_{ v^1 v^{\bar n}}  (q^2) \Big]^2 \frac{q^2}{2x}  \left[ 3 + \frac{q^2}{ m_1^2} + \frac{ (q^2)^2} {m_1^2 s x^2} 
( x - \frac 12)^2 \right]\,.
\label{strutfunct}
\end{eqnarray} 
  
\noindent It is important to remark that $ \bar n , s, q^2  $ and $x$ are not independent, since
\begin{equation}
m_{\bar n}^2 = s = m_1^2 + q^2 ( \frac 1x -1)\,.
\label{condition}
\end{equation}

\noindent 
In the parton model, the structure functions satisfy the Callan-Gross relation 
$F_2 /(2x F_1)=1$. Experimentally this result is confirmed only for certain ranges of $x$ and $q^2$. From eq. (\ref{strutfunct}), we see that the ratio $F_2 /(2x F_1)$ does not depend on the form factors or on the masses of the excited states of the D4-D8 model. This seems to be a consequence of vector meson dominance present in the model. We will investigate numerically this ratio in the next section. 

For the case of axial-vector meson $a_1$, represented by the lowest state $a^1$, we obtain structure functions  with the same form of eq. (\ref{strutfunct}), but with $F_{ v^1 v^{\bar n} } (q^2)$ replace by the axial form factor $F_{ a^1 a^{\bar n} } (q^2)$, defined by 
eq. (\ref{axialformfactors}).

%%%%%%%%%%%%%%%%%%%%%%%%%%%%%%%%%%%%%%%%%%%%%%%%%%%%%%%%%%%%%%%%%%%%

\subsection{Numerical results}

Analyzing the dependence of the calculated vector meson masses $m_n$ with $n$, we perform a quadratic fit, finding
\begin{eqnarray}
m_n^2 \approx -0.0758 - 0.0037n + 0.666 n^2 \,,
\end{eqnarray}
from which we obtain $f(\bar n) = 1/( - 0.0037 + 1.331 {\bar n}) $.

Using the calculated numerical results for the masses and the couplings we can find the structure functions for different values of $x$ and $q^2$. In Figure \ref{F12rhoq2}, we show the behavior of 
$F_1$ and $F_2$ for the $\rho$ meson as functions of $q^2$ with fixed values of the Bjorken parameter $x=0.9, 0.7, 0.5, 0.3$. These functions jump from zero to their maximum values and then fall as $q^2$ increase. A similar behavior is obtained for other values of $x$. We note that as $x$ decreases the maximum value of the structure functions occur for smaller values of $q^2$. 

Note that, eq. (\ref{condition}) implies that, for a fixed value of $x$ there is only a discrete set of $q^2$ values, corresponding to the allowed spectrum of final states $\bar n$. This is a consequence of the fact that we are considering only the contributions from one particle final states. In Figure \ref{F12rhoq2} the plots show discrete sets of points connected by straight lines to illustrate the dependence on $q^2$ for different values of $x$. Note that for larger values of $x$ one has less points in the $q^2$ range considered. 

In Figure \ref{F12rhox}, we show the structure functions for the $\rho$ meson as functions of the Bjorken parameter $x$ for fixed values of $q^2=10,20,30,40{\,\rm GeV}^2$. We note that the maximum of $F_1$ occurs in the region $x\approx 0.8$ and the maximum of $F_2$ occurs in the region $0.8<x<1$. 
Again, eq. (\ref{condition}) implies that, for a fixed value of $q^2$ there is only a discrete set of $x$ values. So the plots in Figure \ref{F12rhox} show discrete sets of points connected by straight lines.   

In Figures \ref{F12a1q2} and  \ref{F12a1x}, we show the structure functions for the $a_1$ meson as functions of $q^2$ and $x$, respectively. The plots and their behavior are similar to the case of the 
$\rho$ meson.

\FIGURE{
\epsfig{file=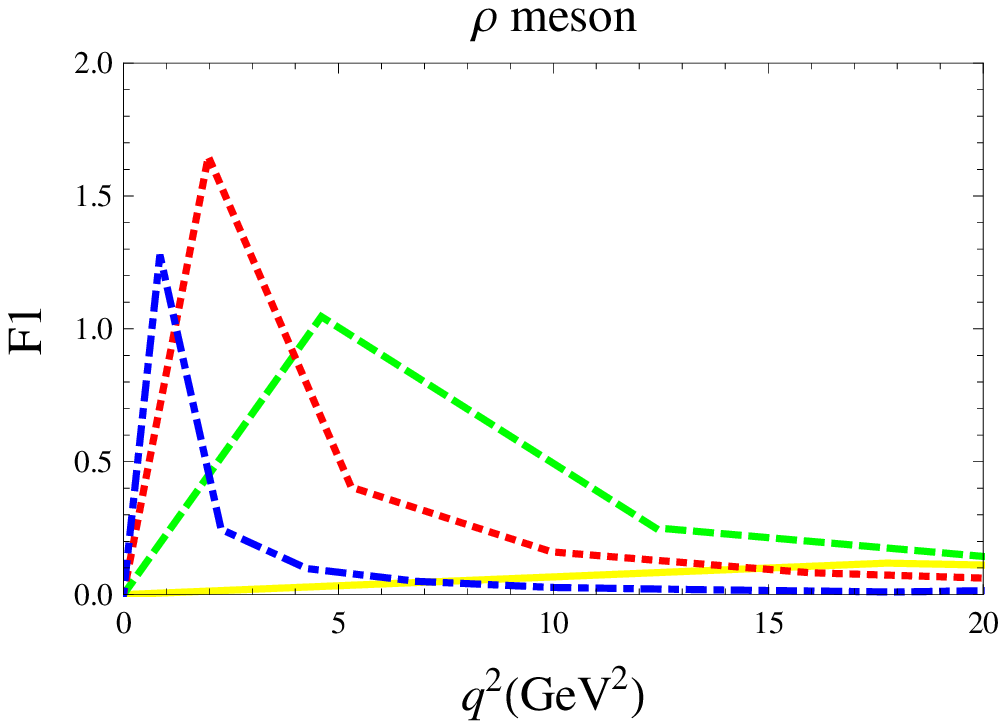, width=7cm} 
\epsfig{file=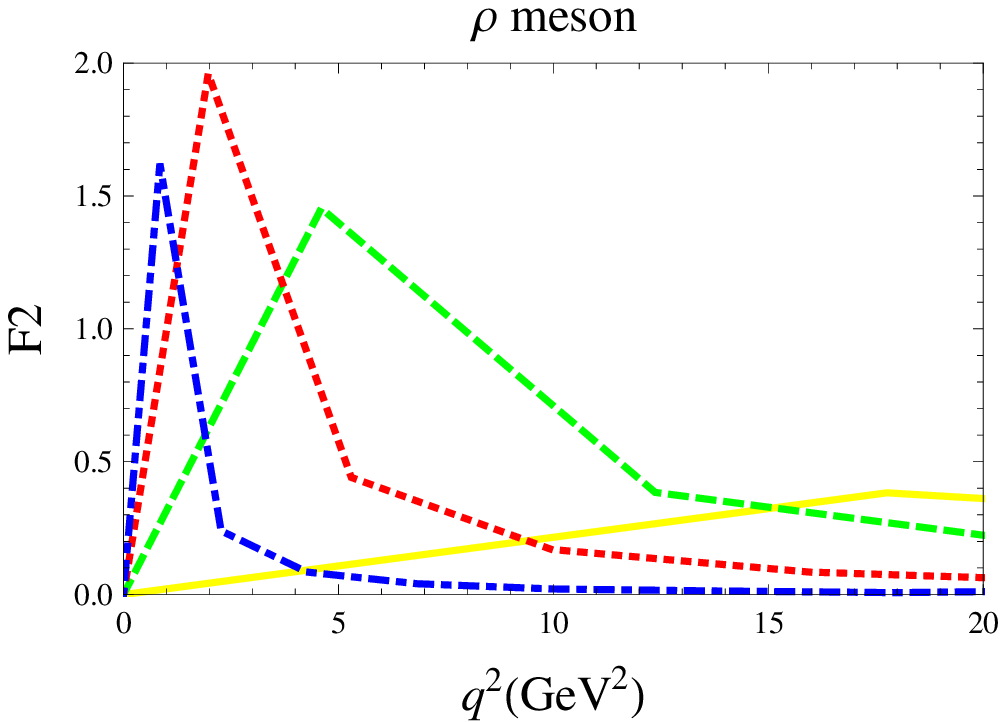, width=7cm} 
\caption{Structure functions $F_1$ and $F_2$ for the $\rho$ meson as functions of $q^2$ for $x=0.9$ (solid yellow line), $x=0.7$ (dashed green line), $x=0.5$ (dotted red line), $x=0.3$ (dot-dashed blue line).} 
\label{F12rhoq2}}

\FIGURE{
\epsfig{file=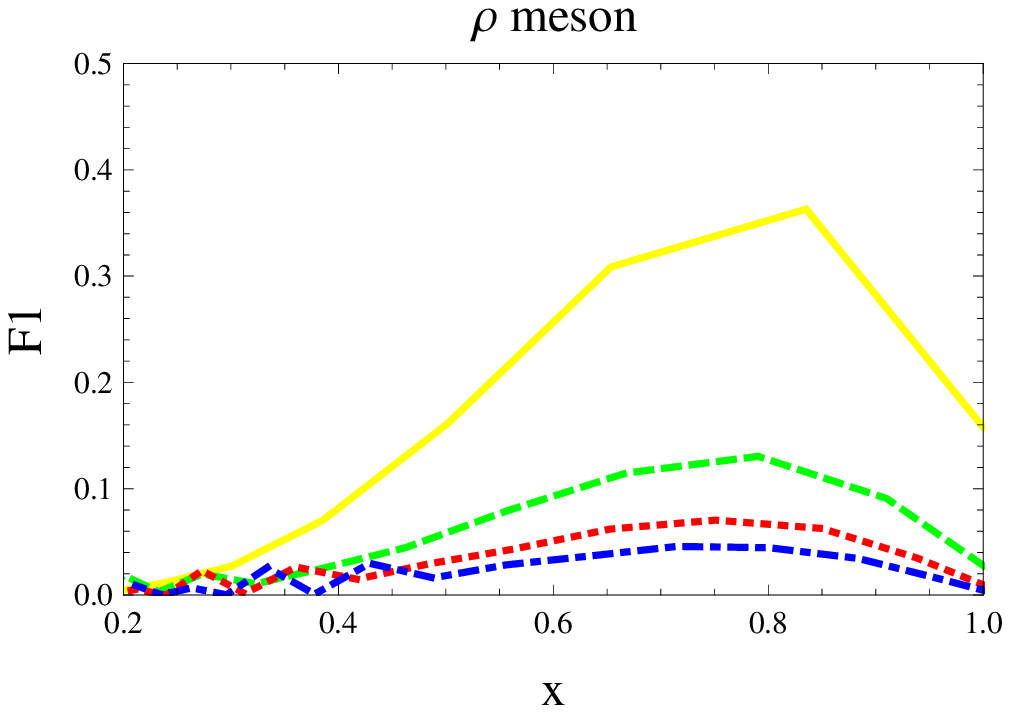, width=7cm} 
\epsfig{file=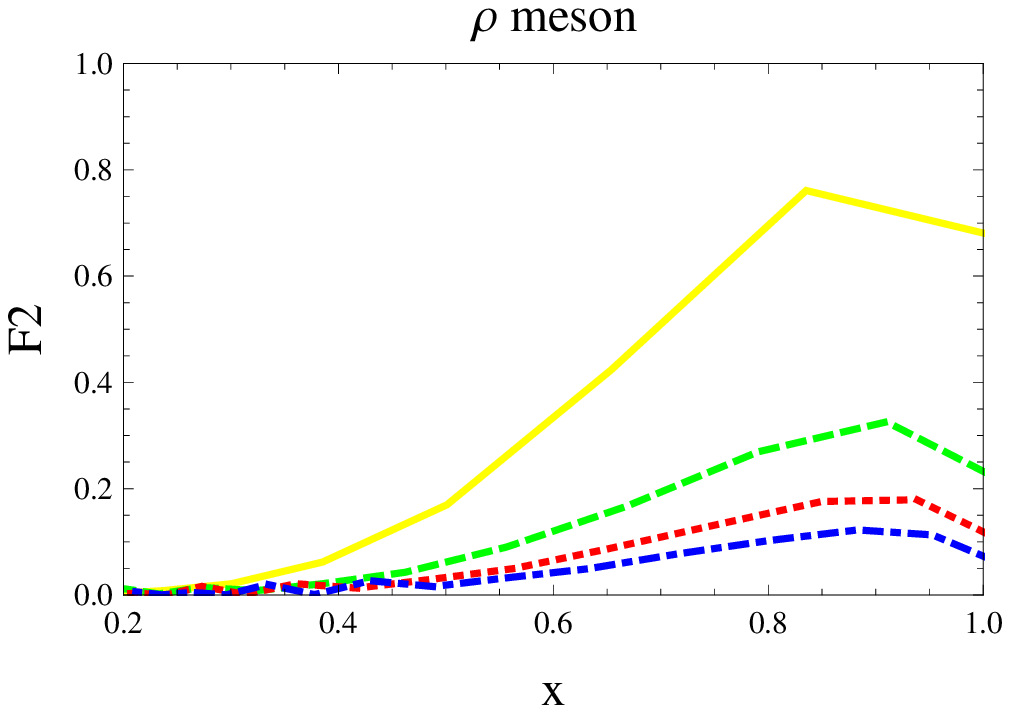, width=7cm} 
\caption{Structure functions $F_1$ and $F_2$ for the $\rho$ meson as functions of $x$ for $q^2=10{\rm GeV}^2$ (solid yellow line), $20{\rm GeV}^2$ (dashed green line), $30{\rm GeV}^2$ (dotted red line), $40{\rm GeV}^2$ (dot-dashed blue line).} 
\label{F12rhox}}

\FIGURE{
\epsfig{file=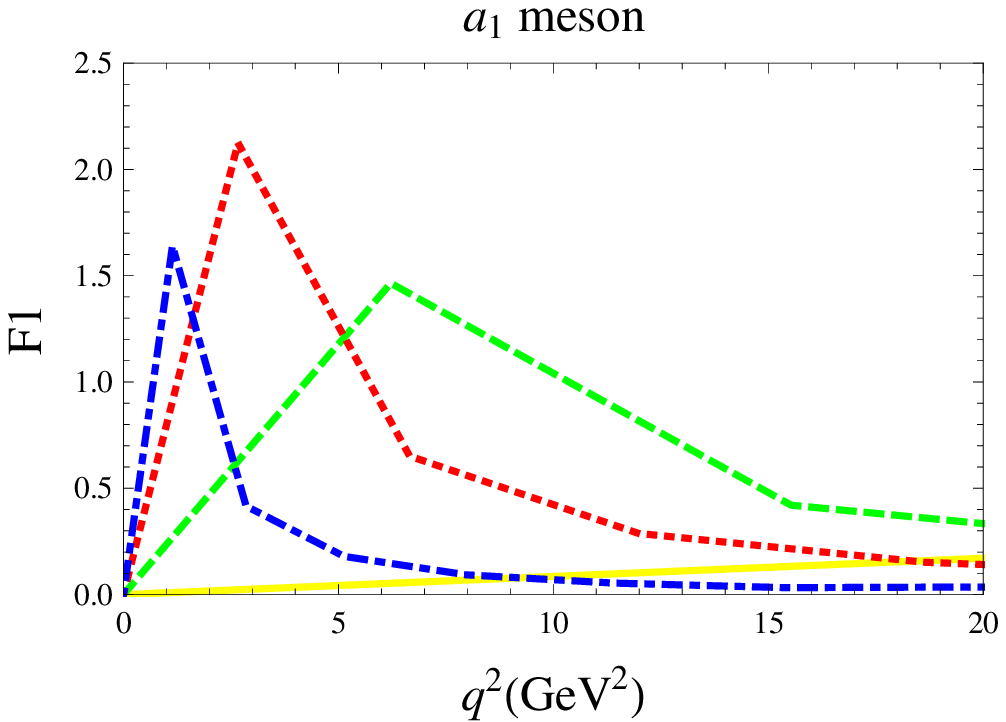, width=7cm} 
\epsfig{file=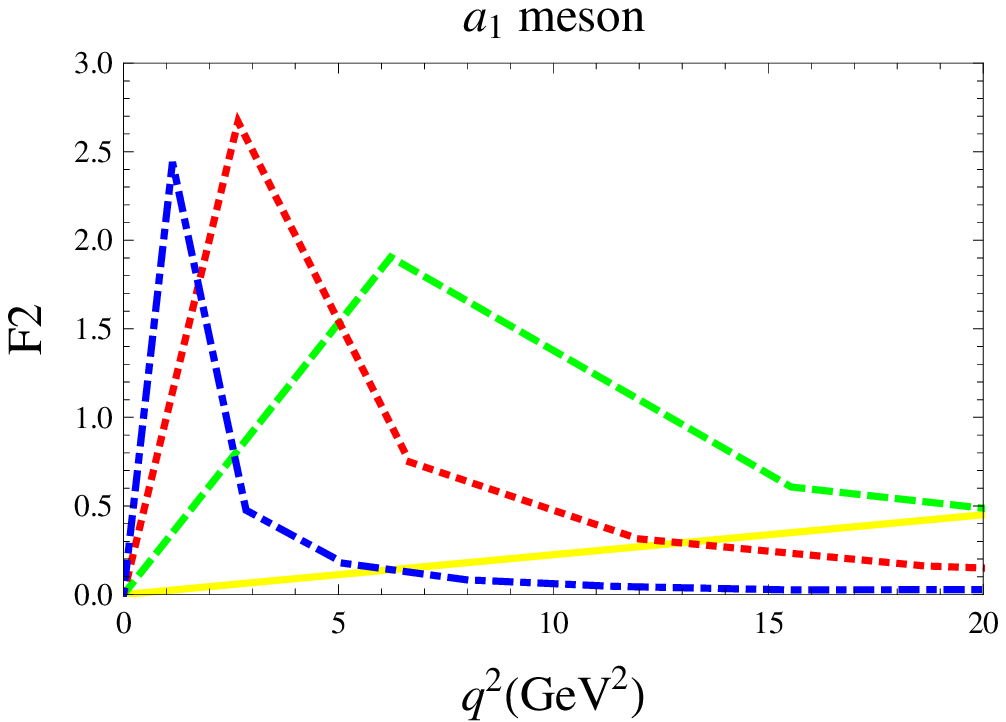, width=7cm} 
\caption{Structure functions $F_1$ and $F_2$ for the $a_1$ meson as functions of $q^2$ for $x=0.9$ (solid yellow line), $x=0.7$ (dashed green line), $x=0.5$ (dotted red line), $x=0.3$ (dot-dashed blue line).} 
\label{F12a1q2}}

\FIGURE{
\epsfig{file=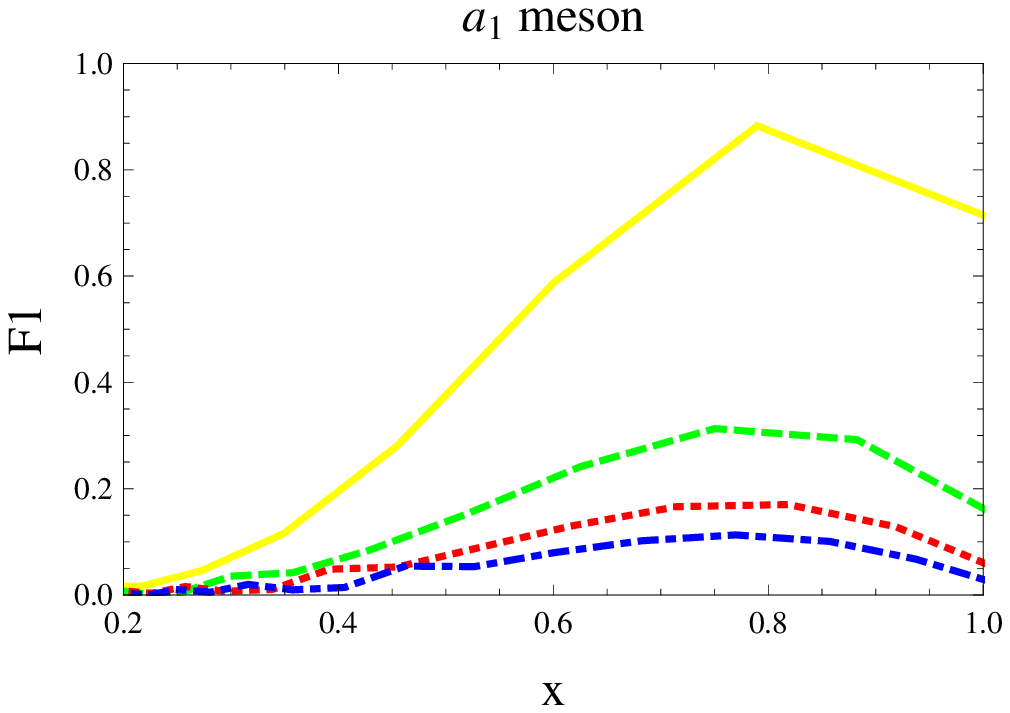, width=7cm} 
\epsfig{file=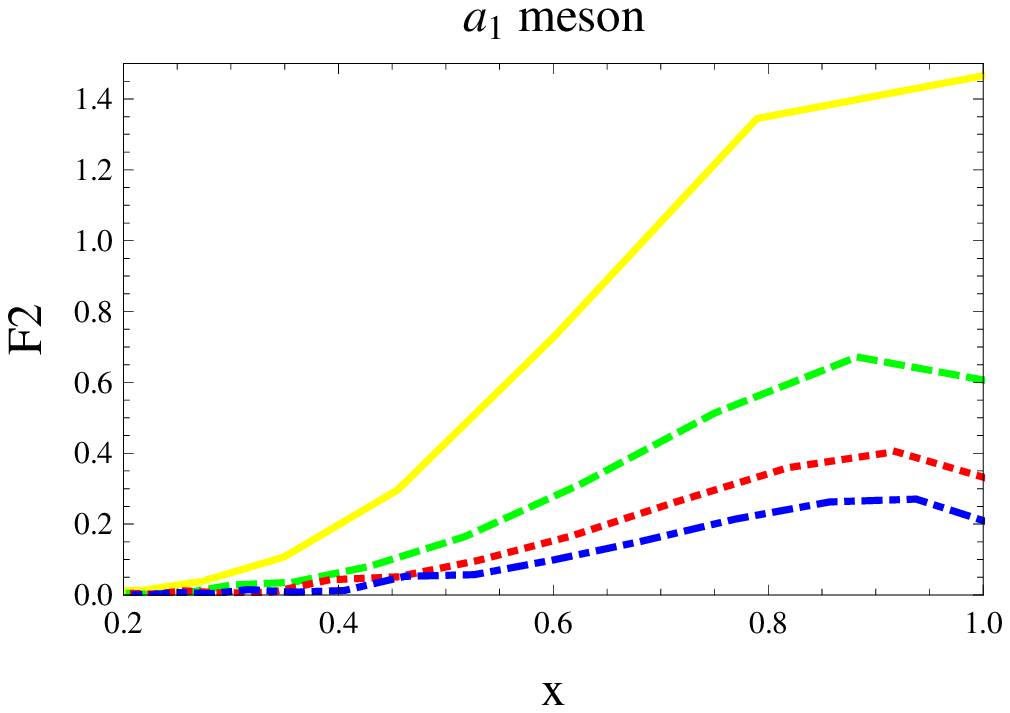, width=7cm} 
\caption{Structure functions $F_1$ and $F_2$ for the $a_1$ meson as functions of $x$ for $q^2=10{\rm GeV}^2$ (solid yellow line), $20{\rm GeV}^2$ (dashed green line), $30{\rm GeV}^2$ (dotted red line), $40{\rm GeV}^2$ (dot-dashed blue line).} 
\label{F12a1x}}

From our numerical results we can study the behavior of the ratio $F_2 /(2x F_1)$ against $x$ and $q^2$. We plot these results in Figures \ref{ComparisonF2F1x} and \ref{ComparisonF2F1q2}. From Figure \ref{ComparisonF2F1x} we note that the Callan-Gross relation $F_2 /(2x F_1)=1$, obtained from the parton model, is verified for all the values of $q^2$ considered in the region 
$0.4 <  x <  0.6$. On the other hand, from Figure 
\ref{ComparisonF2F1q2} we observe that the ratio $F_2 /(2x F_1)$ is approximately independent of $q^2$ for  $q^2>20\,{\rm GeV}^2$ and $x$ fixed in the region $0.3 < x < 0.7$.

\FIGURE{
\epsfig{file=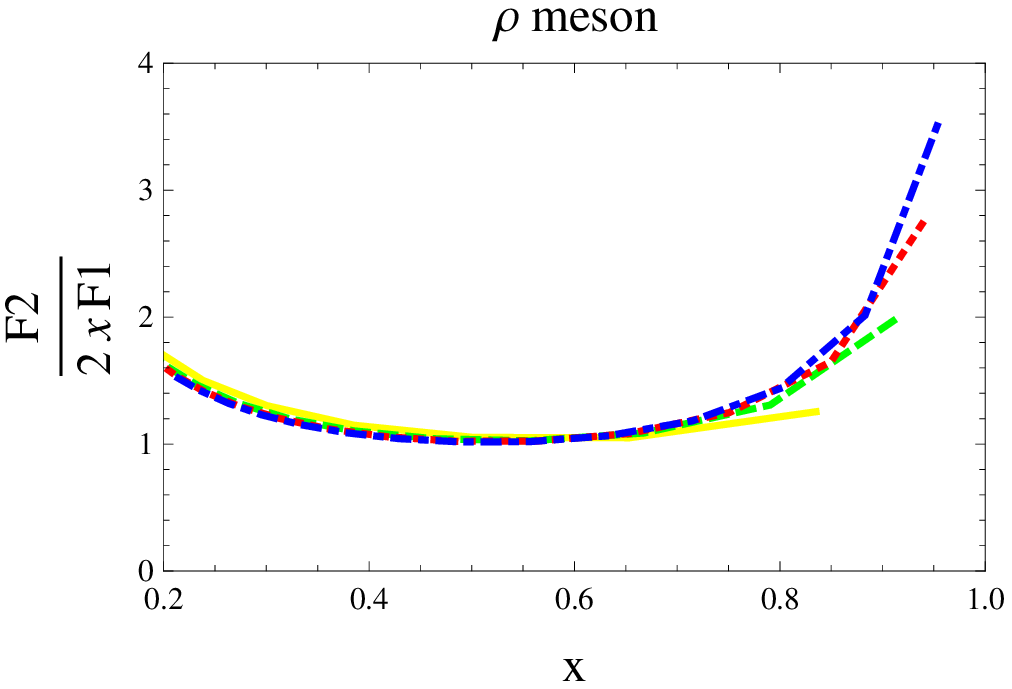, width=7cm}
\epsfig{file=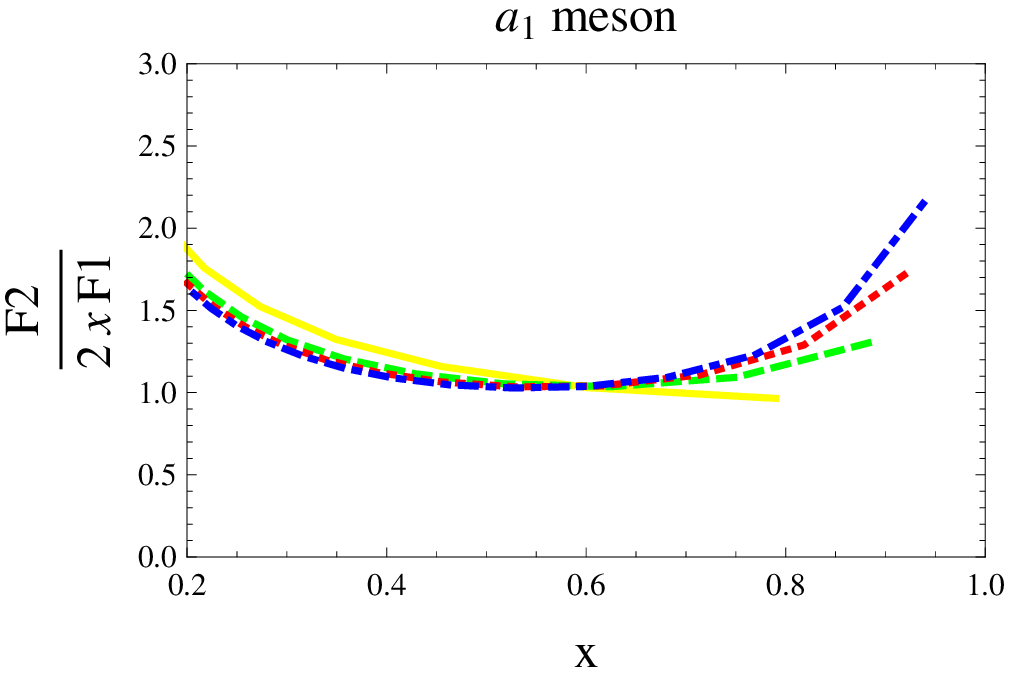, width=7cm} 
\caption{The ratio $F_2 /(2x F_1)$ for the $\rho$ and the $a_1$ mesons as functions of $x$ for $q^2=10{\rm GeV}^2$ (solid yellow line), $20{\rm GeV}^2$ (dashed green line), $30{\rm GeV}^2$ (dotted red line), $40{\rm GeV}^2$ (dot-dashed blue line).} 
\label{ComparisonF2F1x}}

\FIGURE{
\epsfig{file=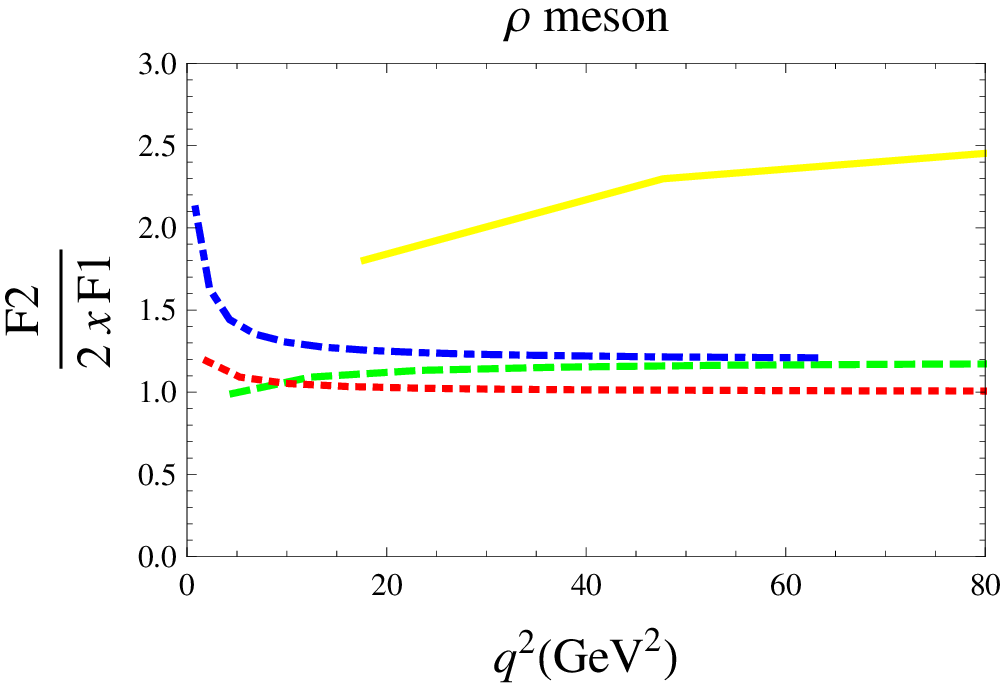, width=7cm} 
\epsfig{file=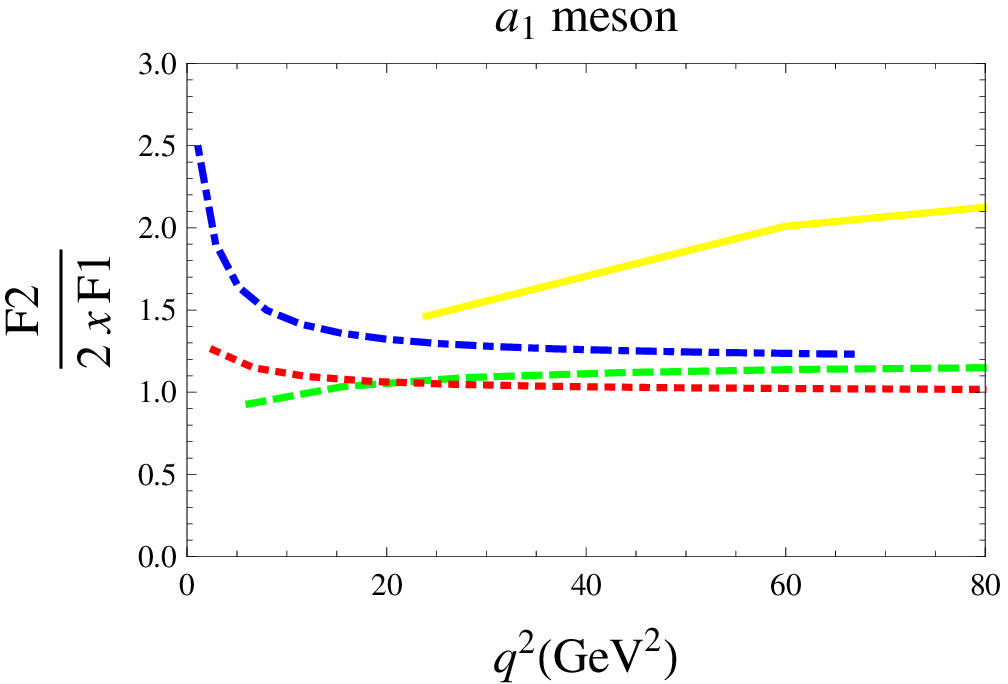, width=7cm}  
\caption{The ratio $F_2 /(2x F_1)$ for the $\rho$ and the $a_1$ mesons as functions of $q^2$ for 
$x=0.9$ (solid yellow line), $x=0.7$ (dashed green line), $x=0.5$ (dotted red line), $x=0.3$ (dot-dashed blue line).} 
\label{ComparisonF2F1q2}}

\section{Conclusion}

In this article we calculated the lowest order contributions to the DIS structure functions for the $\rho$ vector meson and the $a_1$ axial-vector meson. From these structure functions we found that  the ratio $F_2/(2xF_1)$ is approximately equal to one, satisfying the Callan-Gross relation in the interval $0.4 <  x <  0.6$. In the parton model, this relation is a consequence of the spin 1/2 of the hadronic constituents, identified with quarks. 
So, we conclude that even considering only the lowest order contributions to the structure functions, the D4-D8 brane model shows an indication of a quark structure in vector and 
axial-vector mesons. 

We explored here only the structure functions for the lowest states of vector and axial vector mesons but  our results indicate a similar behavior for excited states. Although our analysis  covered the region $ 0.2 < x < 1$ due to the numerical difficulties of the shooting method, it would be interesting to develop other methods to investigate the small $x$ regime and search for gluon saturation and geometric scaling.

\bigskip

\noindent {\bf Acknowledgments:} The authors are partially supported by Capes, CNPq and FAPERJ.

\end{document}